\documentclass[10pt,journal,compsoc]{IEEEtran}

\ifCLASSOPTIONcompsoc
  \usepackage[nocompress]{cite}
\else
  \usepackage{cite}
\fi
\ifCLASSINFOpdf
\else
\fi
\usepackage{amsmath,amssymb}
\usepackage{graphicx}
\usepackage{float}
\renewcommand{\baselinestretch}{1.01}

\begin{document}
\title{Identifying Packet Loss and Reordering Packets\\ in Keyed UDP Transmissions}

\author{F\'abio Machado Gil,
        Nuno M. Garcia, B\'arbara Matos, Nuno Pombo, Rossitza Goleva and~Ciprian Dobre% <-this % stops a space
\IEEEcompsocitemizethanks{\IEEEcompsocthanksitem F\'abio Machado Gil, B\'arbara Matos, Nuno M. Garcia and Nuno Pombo are with the Departamento de Informática, Universidade da Beira Interior, 6200-001 Covilh\~a, Portugal.\protect\\
E-mail: fabio.machado@ubi.pt, anabarbarasmam@gmail.com, ngarcia@ubi.pt, ngpombo@ubi.pt
\IEEEcompsocthanksitem Rossitza Goleva is with the Department of Informatics, New University of Bulgaria, 1618 Sofia, Bulgaria.\protect\\
E-mail: rgoleva@nbu.bg
\IEEEcompsocthanksitem Ciprian Dobre is with the University Politehnica of Bucharest, Bucharest, Romania.\protect\\
E-mail: ciprian.dobre@cs.pub.ro}% <-this % stops a space
\thanks{Manuscript received April 19, 2005; revised August 26, 2015.}}

%{Shell \MakeLowercase{\textit{et al.}}: Identifying Packet Loss and %Reordering Packets in Keyed UDP Transmissions}

\IEEEtitleabstractindextext{%
\begin{abstract}
The User Datagram Protocol (UDP) and other similar protocols send  the application data from the source machine to the destination machine inside segments, without foreseeing nor allowing for any type of control on the transmission or success metrics. These protocols are very convenient for \textit{e.g.} real time data transmission. But when the reliability of the transmitted data is critical, other protocols termed as connection-oriented, allow for full control of the data transmission process, assuring that the received data is an exact copy of the transmitted data, \textit{e.g.} the case of the Transmission Control Protocol (TCP).
To sustain the increased functionality and features of the connection-oriented protocol, a set of mechanisms is implemented based on some specific fields of the segment header. These mechanisms result in a significant overhead in terms of the increased number of transmitted packets. This may further translate into significant delays, because of the additional number of switching and routing tasks, and eventually, because of more complex communications procedures, such as \textit{e.g.} transmission window resizing, and of course, acknowledgement and sequence numbers updating. The two extremes of these communication modalities, one that has no control at all, and the other one that allows for full control, have resulted in the creation of an intermediate protocol that allows for a limited degree of knowledge on how successful a transmission was, and even for an eventual reordering of the segments that arrive out of sequence. This paper presents simulation results that confirm the efficiency of the new almost-reliable UDP protocol, termed Keyed User Datagram Protocol (or KUDP) for transmission of data that includes the ability to identify which packets were lost and to reorder packets that were received out-of-sequence, and points future tasks to be pursued in this research.
\end{abstract}

}

\maketitle

\IEEEdisplaynontitleabstractindextext

\IEEEpeerreviewmaketitle

\ifCLASSOPTIONcompsoc
\IEEEraisesectionheading{\section{Introduction}\label{sec:introduction}}
\else
\section{Introduction}
\label{sec:introduction}
\fi

\IEEEPARstart{T}{he} high overhead that results from the control mechanisms of the Transmission Control Protocol \cite{tcp} (TCP), specifically in terms of the number of packets that need to be transmitted to start, assure the confirmation of successful data reception, and eventually end the transmission, is opposed to the zero overhead that results from the total absence of control messages of the lighter User Datagram Protocol \cite{udp} (UDP). \\

Nevertheless, while data transmitted using TCP is fully controllable, and both the source and the destination have information that allows them to validate the success of the data transmission, or not, with UDP, neither the sender nor the receiver have information regarding the success of the data transmission. This oversimplified introduction omits a large number of other protocols that, using different strategies, are still able to assure a successful data transmission. Yet, the existing protocols can still be considered to be either connection-oriented or connectionless, the first class giving warranty of the data transmission process, and the later trusting that sent packets containing application data will arrive to its destination, which often is not the case.\\

Authors in \cite{kudp} have proposed a new Keyed UDP protocol, in which the use of different port numbers, in a sequence, allow for the destination machine to perceive, to some extent, how successful the transmission was, including, what were the packets that were lost, and also to reorder packets that were received out-of-sequence. While the description of the base concepts for the Keyed UDP are described this recent paper \cite{kudp}, its authors did not provide simulations that proved the advantages of the new protocol. Please note that Keyed UDP does not imply a change in the UDP format \cite{udp} and \cite{kudp} foresees compatibility between applications using UDP and Keyed UDP. \\

This paper presents, in a very condensed manner, the explanation of the basics of the Keyed UDP and the first results obtained by simulation regarding different configuration parameters for the Keyed UDP, including varying loss ratios and ratios for packets received out-of-sequence, and different key lengths for the protocol.\\

As the purpose of this paper is to present the simulation results in first hand, the interested reader may want to refer to \cite{kudp} for a more complete information on the Keyed UDP proposal. Moreover, as this paper only addresses the Keyed UDP, the number of papers cited herein is extremely reduced. Interested readers are also invited to contact the authors to further research on this new topic. Also noteworthy is that although there is a large corpora of science published on UDP, TCP and in general on all protocols, the KUDP proposal is radically new in the sense that uses a sequence of port numbers to transmit data. While all existing protocols send data from one source port to one destination port, KUDP sends data from a sequence of source ports to a sequence of destination ports in a round-robin manner, in its most elaborate configuration. Also, KUDP is not related at all to Multipath TCP (MTCP) \cite{mtcp} as MTCP is supported by several paths using TCP to enhance and render more robust the process of data transmission. A conceptual comparison of KUDP with similar protocols has been already reported in \cite{kudp}.\\

The remainder of this paper is organized as follows: this paragraph concludes Section \ref{sec:introduction}; Section \ref{sc:keyedudp} presents a quick overview of the Keyed UDP; Section \ref{sc:Simulations} presents the design of the simulations and the obtained results; Section \ref{sc:Discussion_and_Conclusions} concludes the paper, including the discussion of the results, the conclusions and future steps.

\section{Keyed UDP}\label{sc:keyedudp}
\begin{figure}[h]
 \centering
 \includegraphics[width=8cm]{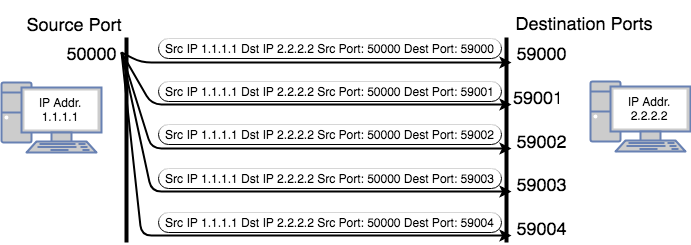}
 \caption{Keyed UDP transmission (adapted from \cite{kudp}).}
 \label{fig:Image1}
\end{figure}

\begin{figure}[h]
 \centering
 \includegraphics[width=8cm]{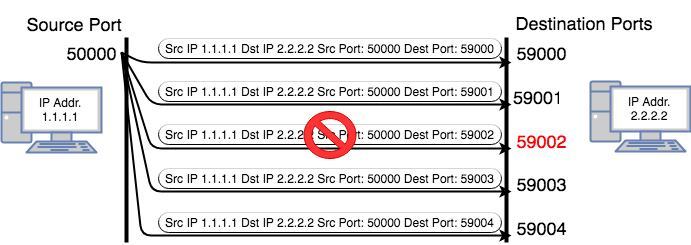}
 \caption{Keyed UDP transmission with one packet lost (adapted from \cite{kudp}).}
 \label{fig:Image2}
\end{figure}

While UDP, TCP and other protocols send segments from one port in the source machine to another port in the destination machine, the Keyed UDP protocol sends packets, for example, from one port in the sender machine to a series of ports in the destination machine. The receiver machine opens a set of ports and expects the packets to be received in a given order, \textit{e.g.} the sender sends the packet using its IP address and its port, say, port 50000, and sends the packets to the destination IP address, and to ports \textit{e.g.} 59000, 59001, 59002, 59003, 59004. After receiving the packets, the destination machine can forward its payload to the relevant application. Figures \ref{fig:Image1} and \ref{fig:Image2} show a visualization of a data transmission where all packets were received successfully, and another where one of the packets was lost.

Yet, if the received packets are received in a different order, say, 59000, 59001, 59003, 59002, 59004, then the destination machine can reorder the packets based on the expected sequence for the port number. Or, if packets arrive at ports 59000, 59001, 59003 and 59004, the destination machine can infer that one packet was lost (depicted in Fig. \ref{fig:Image2}). This trivial example considered a key of length 5, being the key length the number of ports minus one used for the data transmission. It also considers that the ports were open in the destination machine, on what authors in \cite{kudp} termed destination keyed UDP (dKUDP). Keys can be applied at the sender side, or in both the sender and the receiver side, keys can be non-sequential and keys can be dynamic over time, as authors in \cite{kudp} propose.\\

For the sake of simplification of this summary explanation on the possible working of the Keyed UDP, it shall be assumed that both the source and the destination applications have hardcoded information on the ports that are used during data transmission, much in a similar manner to a HTTP server that "knows" that it has to listen to port 80, and a HTTP client "knows" that it needs to address its requests to port 80 of the server. Nevertheless, authors of \cite{kudp} considered other means of exchanging or inferring the information relative to the communication key.\\

In \cite{kudp} an algorithm to recover an out-of-sequence and lossy sequence of packets was also proposed and termed Stream Reconstruction Algorithm (SRA for short). This algorithm uses an election process on a list of multiple iterations on an array of received packets. While other algorithms may be developed with that purpose, this research used the proposed election algorithm in the simulations.\\

Authors of the Keyed UDP do not consider what happens at the destination machine after a packet was discovered missing, and leave the actions that may be taken open for additional contributions. The idea that the Keyed UDP can to some extent identify which packets are missing, leaves room to, for example, request them again from the sender machine, or in the case of real time communication, to use the data received in the precedent and subsequent packets to infer the content of the missing packet, by performing data imputation. In \cite{kudp} the authors present a number of open issues, being the ones interesting to this research as follows:
\begin{enumerate}
	\item the simulation and validation of the efficiency of Keyed UDP;
	\item the trade-off between the number of ports in a key and the effort posed to the Stream Reconstruction Algorithm.
\end{enumerate}

In particular, point 2) refers to another issue that was not addressed in \cite{kudp}, this being the efficiency of the protocol and the performance of SRA with different key lengths. The following section describes how the simulations were designed to address this question, and in conclusion, allow the assessment of the usefulness of the Keyed UDP.\\

Regarding the computational cost for the operating systems of having a large number of open ports, either for sending and/or for receiving data, this consists mostly, in the classical approach of the client-server model, in the addition of new variables, usually for port or communication handles, to allow the reception of the data forwarded from the TCP/IP protocol stack that is implemented. This scenario is similar to opening a new tab in a browser application and making a new web page request, being therefore quite common. Authors in \cite{kudp} considered the alternative of a new design for the TCP/IP protocol stack, in which case, the segments would be handed to a transport layer operating in promiscuous mode, that would receive the communications to all ports, processing it as adequate.\\

KUDP may be used in scenarios where data transmission is important but not critical, such as \textit{e.g.} in sensor networks, where a sensor needs to transmit a small amount of data, but transmitting it at a particular time or at a later time is not relevant for the solution, or \textit{e.g.} in applications for real time voice and/or video communication where data imputation of a lost parcel of data is feasible. The use of KUDP as a substitute of TCP is not foreseen at this stage. Finally, to the authors' best knowledge, this is the only protocol that allows for some degree of information regarding the success and out-of-sequence correction of received packets using an UDP segment format with transmission zero-overhead.

\section{Simulation and results}\label{sc:Simulations}
To assess the efficiency of the protocol, a number of tests were devised. Generally the tests can be described as follows:
\begin{enumerate}
    \item Generate a number of packets that will be sent to a destination machine, creating the initial sequence;
    \item Using a predefined threshold for loss or for packets received out-of-sequence, determine if a given packet is going to be lost or going to be received out-of-sequence; if the packet is marked to be received out-of-sequence, swap it with the subsequent packet;
    \item After the initial sequence has been processed containing missing packets or packets received out-of-sequence, the destination machine applies the SRA, creating the reconstructed stream;
    \item Compare the reconstructed stream with the initial stream and compute the errors, these being the number of missing packets not correctly identified, or the number of packets received out-of-sequence not correctly placed in order.
\end{enumerate}

The computation of the efficiency is also here interpreted in a strict sense, \textit{i.e.}, the success of the algorithm is accounted for, only if the packet in the final sequence is at the exact same position in the initial sequence, or if the packet was lost and the its loss was duly identified. Therefore, the case of false negatives and the case of false positives are both considered as errors.\\ 

These tests considered only dKUDP, and each simulation for this test was run 100 times. The results shown in this section are the average of all the runs. Also, as the probability for a packet loss or out-of-sequence packet was random and calculated for each packet, so the overall ratio for induced errors in the transmission is always a number near the predefined threshold. \\
The parameters for the simulation were defined as follows:
\begin{itemize}
    \item Number of generated packets per stream = 4 times the size of the key
    \item Number of runs for each simulation = 100
    \item Key lengths considered = 5, 10, 15, 20, 50, 100 and 200
    \item Thresholds for packet loss = from 1\% up to 25\%
    \item Thresholds for packet received out-of-sequence = from 1\% up to 70\%
\end{itemize}

In this simulation the number of generated packets that compose the initial stream is limited to four times the size of the key, particularly because as the SRA is intended to be used in real time, with the election occurring as the packets arrive, the effect of the SRA is limited to a range that is never larger than the size of the key. Yet this is not a real limitation as each simulation was run 100 times and the results here shown are the average of all simulations for each scenario. Also, the lower thresholds for lost packets was set to 25\% for two reasons: 
\begin{enumerate}
    \item the computation of the efficiency metric of the algorithm is strict, \textit{i.e.}, if a packet was expected to arrive in the \textit{n\textsuperscript{th}}, and instead arrives at the \textit{(n+k)\textsuperscript{th}} position, it is still considered an error. Yet, if a sequence of \textit{k} packets is lost, this event should not be interpreted as an error as long as the \textit{k} missing packets are acknowledged for;
    \item the goal of this initial simulation was to assess the higher boundary of success of the algorithm, as there are yet no criteria that allow the lower boundary of efficiency where the quality of the communication \textit{e.g.} of real time multimedia communication is still acceptable in terms of user experience.
\end{enumerate} 

Nevertheless, this research has some limitations. First and foremost, the joint occurrence of events of loss and out-of-sequence of packets was not simulated and will be object of additional research. Additionally, the SRA described in \cite{kudp} considered a size for the buffer equal to \textit{n-1}, where \textit{n} is the size of the key. This research also experimented with varying the size of the SRA buffer as a function of the size of the key, and the reported results here consider the buffer size to be half the size of the key, rounded by excess.\\

A final limitation of this research is the absence of a method to identify the end of the transmission. Authors in \cite{kudp} refer that as an open issue, suggesting that this can be detected using a timer. As this is relevant to the simulation because the initial and the final elections are the least populated, as there are not as many packets as in the middle of the transmission to allow the election process, the simulation assumes that the stream of packets is ended as suggested, by means of a time-out.\\

The SRA was implemented as suggested in \cite{kudp}. Considering \textit{k} as the length of the key, the basic rules of operation can be described as follows:
\begin{enumerate}
    \item The first set of packets of length \textit{k} is received;
    \item The packets are sorted into place considering its destination port number; a list is created to hold the sorted packets; 
    \item eventually some positions of this list will be empty, and these are marked as \textit{f} for "failure to receive";
    \item The election process takes place; this is performed as follows:
    \begin{enumerate}
        \item The first packet can be elected after the first list is created (when in fact there is only one candidate for the first position);
        \item While the number of packets received is less than \textit{k}, the election is done with a smaller list of candidates;
        \item The \textit{n\textsuperscript{th}} packet can be elected after the \textit{n\textsuperscript{th}} list is created;   
    \end{enumerate}
    \item If a new packet arrives, and a new list is created, repeating points 2, 3 and 4;
    \item Else, if no more packets arrive, finish the elections repeating point 4 until all received packets are elected.
\end{enumerate}

Authors in \cite{kudp} provide an example of the election process. 
The implemented election process follows these rules:
\begin{itemize}
    \item For a given position, the candidate with most occurrences wins;
    \item In case of a tie, the candidate with the first occurrence wins;
    \item If a packet is received and never elected, it will be placed in the first available suitable position (hidden candidate);
    \item Once elected, a packet can not be candidate to a new position.
\end{itemize}

\begin{figure}[h]
 \centering
 \includegraphics[width=8cm]{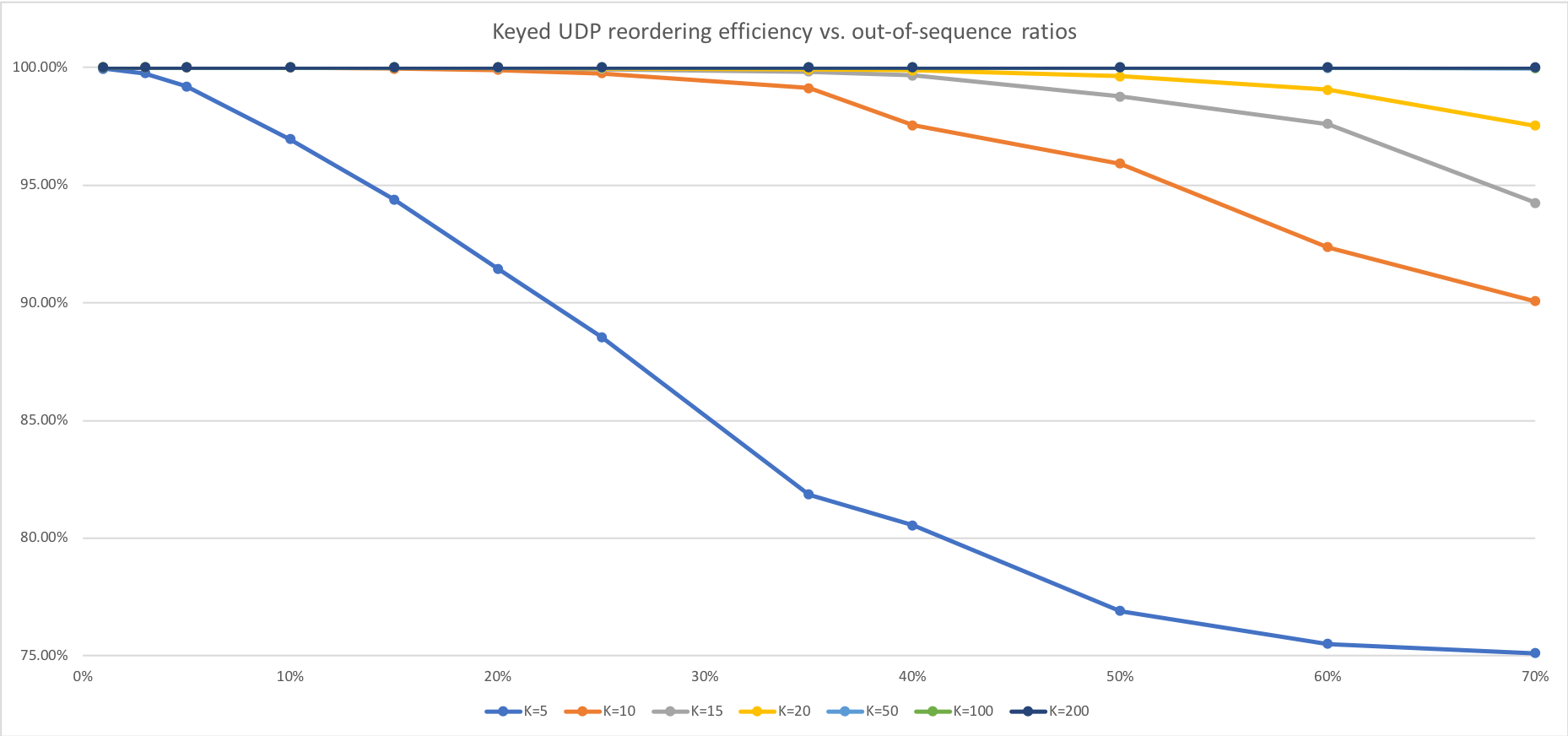}
 \caption{Efficiency of the SRA for different packets swapping ratios and different key lengths.}
 \label{fig:Image3}
\end{figure}

\begin{figure}[h]
 \centering
 \includegraphics[width=8cm]{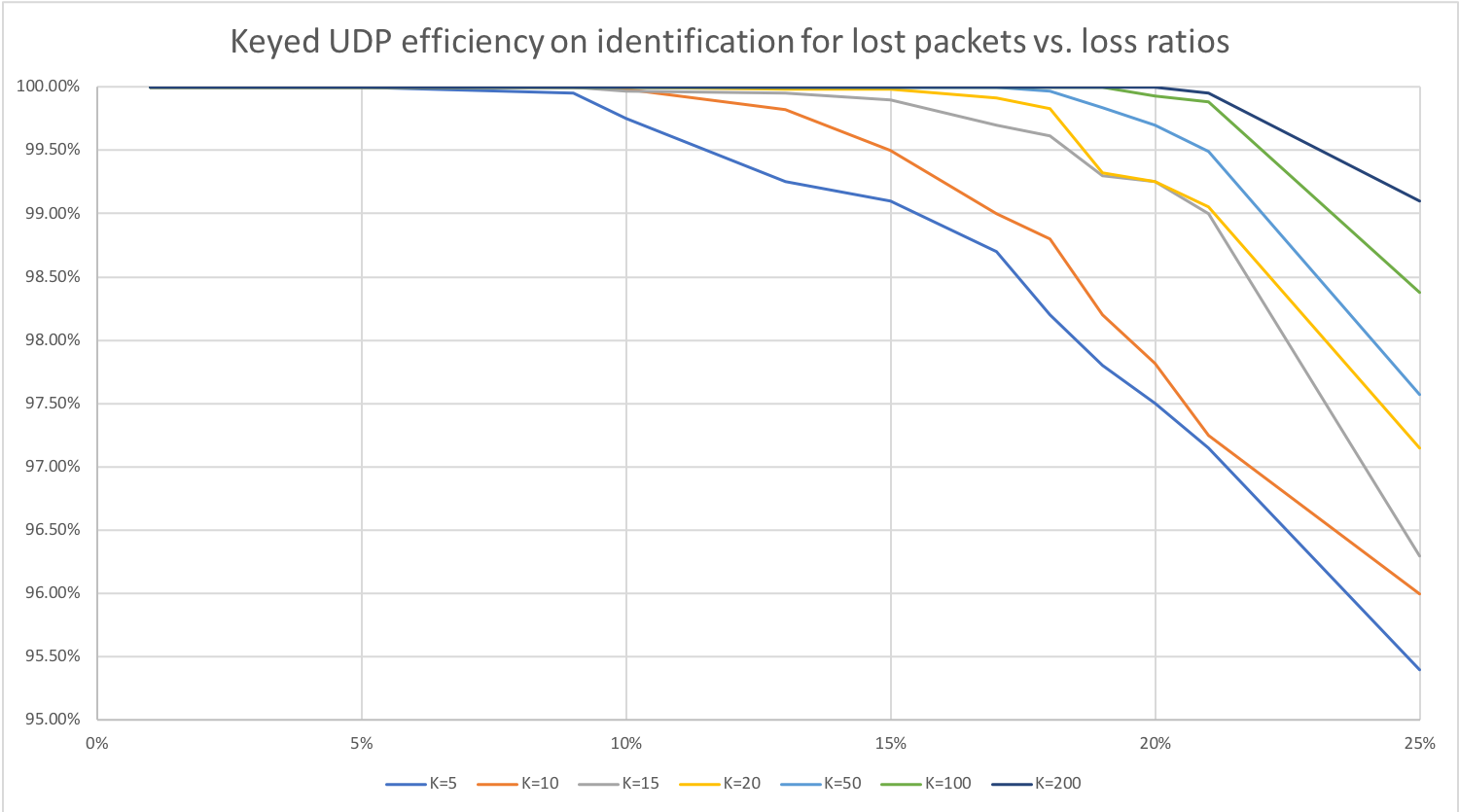}
 \caption{Efficiency of the SRA for different packets loss ratios and different key lengths.}
 \label{fig:Image4}
\end{figure}

The results described in Tables \ref{tab:Table1} and \ref{tab:Table2} are shown in a graphical manner in Fig. \ref{fig:Image3} and Fig. \ref{fig:Image4}. Fig. \ref{fig:Image3} shows the efficiency ratio of the SRA for different packet swapping ratios. The ratio of packets that will arrive out-of-sequence is a parameter in the simulation, which develops as follows: when a packet is generated, a random number is launched to determine if that packet is to be swapped with the subsequent packet or not. If the packet is to be swapped, then it's placed in the next position of the received array. Yet, if the subsequent packet is also market for swapping, it will occupy the next position, thus making both the previous and the current swapping operations cancel each other.\\ 

One of the expected results for this simulation if that for very large swap ratios, a significant number of packets gets swapped and eventually, they end up composing a correct sequence again, justifying the inversion of the concavity of the curve that can be observed with greater evidence in the curve for key length of 5 (\textit{k=5}). \\

Also as expected, for low packet swapping ratios, longer key length allows the SRA to recover from 100\% packet swaps. For a key length of 5 ports, the shortest simulated key, and for packet swapping ratios from 1\% to 70\%, the SRA allows for the reconstruction of 99.95\% of the original payload down to 75.10\%. Table \ref{tab:Table1} shows some of the most relevant values for this feature.

\begin{table}[h]
\label{tab:Table1}
\caption {Efficiency of SRA for packets received out-of-sequence \textit{vs.} key length.}
\centering
\begin{tabular}{r|r|r|r|r|r|}
\cline{2-6}
                                     & \textbf{10\%} & \textbf{35\%} & \textbf{50\%} & \textbf{60\%} & \textbf{70\%} \\ \hline
\multicolumn{1}{|r|}{\textbf{K=5}}   & 96,95\%       & 81,85\%       & 76,90\%       & 75,50\%       & 75,10\%       \\ \hline
\multicolumn{1}{|r|}{\textbf{K=10}}  & 100,00\%      & 99,12\%       & 95,92\%       & 92,37\%       & 90,07\%       \\ \hline
\multicolumn{1}{|r|}{\textbf{K=15}}  & 100,00\%      & 99,83\%       & 98,76\%       & 97,61\%       & 94,24\%       \\ \hline
\multicolumn{1}{|r|}{\textbf{K=20}}  & 100,00\%      & 99,92\%       & 99,63\%       & 99,06\%       & 97,53\%       \\ \hline
\multicolumn{1}{|r|}{\textbf{K=50}}  & 100,00\%      & 100,00\%      & 100,00\%      & 99,99\%       & 99,97\%       \\ \hline
\multicolumn{1}{|r|}{\textbf{K=100}} & 100,00\%      & 100,00\%      & 100,00\%      & 100,00\%      & 99,99\%       \\ \hline
\multicolumn{1}{|r|}{\textbf{K=200}} & 100,00\%      & 100,00\%      & 100,00\%      & 100,00\%      & 100,00\%      \\ \hline
\end{tabular}
\end{table}

\begin{table}[H]
\label{tab:Table2}
\caption {Efficiency of SRA for packets lost \textit{vs.} key length.}
\centering
\begin{tabular}{r|r|r|r|r|}
\cline{2-5}
                                     & \textbf{10\%} & \textbf{15\%} & \textbf{20\%} & \textbf{25\%} \\ \hline
\multicolumn{1}{|r|}{\textbf{K=5}}   & 99,75\%       & 99,10\%       & 97,50\%       & 95,40\%       \\ \hline
\multicolumn{1}{|r|}{\textbf{K=10}}  & 99,98\%       & 99,50\%       & 97,82\%       & 96,00\%       \\ \hline
\multicolumn{1}{|r|}{\textbf{K=15}}  & 99,97\%       & 99,90\%       & 99,25\%       & 96,30\%       \\ \hline
\multicolumn{1}{|r|}{\textbf{K=20}}  & 100,00\%      & 99,98\%       & 99,25\%       & 97,15\%       \\ \hline
\multicolumn{1}{|r|}{\textbf{K=50}}  & 100,00\%      & 100,00\%      & 99,70\%       & 97,57\%       \\ \hline
\multicolumn{1}{|r|}{\textbf{K=100}} & 100,00\%      & 100,00\%      & 99,93\%       & 98,38\%       \\ \hline
\multicolumn{1}{|r|}{\textbf{K=200}} & 100,00\%      & 100,00\%      & 100,00\%      & 99,10\%       \\ \hline
\end{tabular}
\end{table}

Table \ref{tab:Table2} shows some of the most relevant values for the packet loss scenario. The simulated values for packet loss ratio varied from 1\% to 25\% (not all shown in the table). It can be seen that the identification of the missing packets is in average achieved for over 95\% of the packets independently of the key size and of the packet loss ratio. This means that the SRA was able to identify more than 95\% of the packets that were lost even when 25\% of the packets were lost. This performance peaks 99.10\% of lost packet identification when a key of 200 ports is used.\\

Fig. \ref{fig:Image4} shows the plot data for the efficiency of SRA when the stream of packets is subject to packet loss. As expected, longer keys return higher efficiencies, \textit{i.e.}, Keyed UDP using longer keys (100 ports or more) is able to identify more than 98.38\% of the packets that were lost at a 25\% packet loss ratio. Also, at this loss ratio, the 50 ports long key enables Keyed UDP to identify 97.57\% of the packets that were lost, and at a 15\% loss rate, the 50 ports key is also able to identify 100\% of the lost packets.\\

\section{Discussion and Conclusions}\label{sc:Discussion_and_Conclusions}
The User Datagram Protocol is used when fast data transmission is necessary and data transmission success control can or must be waived, as this protocol does not provide any type of control and therefore, does not allow for any success metric. On the contrary, and at the expense of communication control mechanisms implemented by control packets, the Transmission Control Protocol allows for a fully controllable data transmission, including lost packet re-transmission through the use of sequence and acknowledgment numbers that both the sender and the receiver exchange.\\

The room for an almost reliable protocol seems to exist in a number of situations, first and foremost in multimedia real time communications, where the receiving machine can use the ability to compute transmission success metrics that may allow for additional features. Other possibility also described in the Keyed UDP initial paper considers the use of this protocol in Machine-to-Machine communications, when data integrity is relevant, but if a corrupted communication is received, the system is not compromised and is able to recover in the next transmission slot.\\

The Keyed UDP proposal, as described in \cite{kudp}, maintaining the format of the UDP segment, and promising a zero overhead data transmission with some degree of control, was further researched and its efficiency was assessed using simulation. The scenarios of packet loss and packets received out-of-sequence were simulated, for different levels of severity of these transmission hazards, and for different key lengths.\\

As expected, longer keys return higher efficiency results, and for example, a key with length of 50 ports can identify and correct 99.97\% of the packets received out-of-sequence, and for keys of 100 ports, if a stream is received having 70\% of the packets that were received out-of-sequence, the SRA can reconstruct the original stream up to 99.99\%.\\

Also in the scenario of packet loss, a key of 50 ports can identify which packets were lost with 97.57\% accuracy, considering a 25\% loss ratio. For a key with 200 ports and 25\% lost packets, the success ratio of the reconstruction algorithm can reach 99.10\%.\\

As main conclusions, it can be noted that the implemented SRA allows for the demonstration of the success of the almost-reliable Keyed UDP protocol, given the rates of success for the reconstruction of the original sequences with either packets received out-of-sequence and with packets lost.\\

Another conclusion may be that, given the degree of success in the identification of the lost packets, larger keys need to be researched, and the SRA needs to be further fine tuned. For example, it would be very interesting to see what happens when the size of the SRA buffer is not a linear function of the size of the key, mostly because the size of the SRA buffer has impact in the delay of the data being handed over to the application layer.\\

Also as additional research, the combined occurrence of packets received out-of-sequence and packets lost needs to be addressed, keeping the ratios for these two types of events at a realistic level. Unfortunately, it was not possible to search the literature for realistic levels of packets arriving out-of-sequence not for packets that were lost.\\

Finally, it is also believed that the length of the simulations need to be defined not in function of the length of the key, but for a predefined set of numbers of transmitted packets.\\

\ifCLASSOPTIONcompsoc
  \section*{Acknowledgments}
\else
  \section*{Acknowledgment}
\fi
This work is funded by FCT/MEC through national funds and when applicable co-funded by FEDER – PT2020 partnership agreement under the project \textbf{UID/EEA/50008/2019} (\textit{Este trabalho é financiado pela FCT/MEC através de fundos nacionais e quando aplicável cofinanciado pelo FEDER, no âmbito do Acordo de Parceria PT2020 no âmbito do projeto UID/EEA/50008/2019}).

This article is based upon work from COST Action IC1303 - AAPELE - Architectures, Algorithms and Protocols for Enhanced Living Environments and COST Action CA16226 - SHELD-ON - Indoor living space improvement: Smart Habitat for the Elderly, supported by COST (European Cooperation in Science and Technology). More information in www.cost.eu.

\ifCLASSOPTIONcaptionsoff
  \newpage
\fi



%\bibliographystyle{IEEEtran}

\end{document}